\documentclass[12pt,article]{article}
\usepackage{fancyhdr}
\usepackage{float}

\usepackage{amsmath,amssymb}
\usepackage{physics}
\usepackage{wrapfig,lipsum}
\usepackage{mathtools}
\usepackage{graphicx}
\usepackage{color}
\usepackage{multicol}

\begin{document}

  \date{}
  
  \title{\Large\bf Flooding simulation due to Hurricane Florence in North Carolina with HEC RAS}
  

  \author{ 
    Alvin Peng\textsuperscript{1},  Fei Liu\textsuperscript{2} \\
    \textsuperscript{1} \textit{South Brunswick High School} \\
    \textsuperscript{2} \textit{New Jersey Science Academy, Fei.Liu@njsci.org}
    }

  \maketitle

  \section*{\centering Abstract}
  Flooding due to Hurricane Florence led to billions of dollars in damage and nearly a hundred deaths~\cite{Fox19} in North Carolina. These damages and fatalities can be avoided with proper prevention and preparation. Modelling such flooding events can provide insight and precaution based on principles of fluid dynamics and GIS technology. Using topography and other geographic data from USGS~\cite{USGS}, HEC-RAS  ~\cite{HR19A} ~\cite{HR19B} can solve the Shallow Water Equations over  flooding areas to assist the study of inundation patterns. Simulation results from HEC-RAS agree with observations from NOAA ~\cite{NOAA_FL1} ~\cite{NOAA_FL2} in the flooding area studied. Modeled results from specific locations affected by Hurricane Florence near Neuse River, NC are compared with observations. While overall pattern of inundation is agreeable between model results and observations, there are also differences at very specific locations. Higher resolution topography data and precipitation data over the flooding area may improve the simulation result and reduce the differences.

  \textit{}
  
  
  \section{Introduction}
  
The origin of the word “flood” comes from the Old English word fl\={o}d. It shares the same roots as the words “flow” and “float,” two terms associated with water. A flood is when there is a substantial overflow of water into usually dry land. There are three common types of flooding: fluvial, pluvial, and coastal. A fluvial flood, or a river flood, is used to describe when the water levels of bodies of water such as rivers and streams rise and spill out onto the nearby banks. Water level rise could be attributed to rain or snowmelt. River flooding probability can be determined based on past precipitation data, river levels, soil type and terrain. Hurricane Florence caused high amounts of fluvial flooding. Pluvial floods occur in areas without a nearby body of water. Usually, pluvial flooding occurs when large amounts of rain overwhelms drainage systems. Coastal flooding occurs along coastal areas near seas and oceans. Intense storms, storm surges and tsunamis can push seawater into low-lying coastal land. The size, speed and direction of a storm, as well as topography, determine the severity of coastal floods.

Floods have socioeconomic and environmental consequences. The short term negative consequences of floods such as the loss of human life and property are the most noticeable. Damages to buildings and other structures are immediate and obvious. Floods can damage key infrastructure not limited to but including roadways, communication links and power plants. Damage to roadways and communication can make it difficult for first responders to get to people in need of help. Damage to power lines can disable hospitals and water treatment facilities. Additionally, flood waters carrying sewage or chemical waste, and a lack of clean water increase risk of disease. Floods may leave certain farmlands unworkable, causing food shortages. Rebuilding can take months or years depending on the severity of the damage, and can disrupt everyday life. Roads may get blocked and people may be unable to go to work or school. Long-term economic setbacks caused by flooding include a decrease in tourism and a loss in property value. 40 percent of small businesses never reopen after a flood disaster~\cite{FEMA}. In some cases, extremely flood-prone areas may be deemed uninhabitable and relocation may be necessary. As for the long-term effects on human health, wet buildings from floods can lead to indoor mold growth which has been linked to some respiratory problems, and psychological damages and injuries may stay with people for the rest of their lives. 

On the other hand, smaller floods that are not in urban areas can be beneficial to the ecosystems. Floods can replenish groundwater. In certain dry geographical regions, agriculture relies on seasonal floods to bring water and nutrients to soils and to kill pests. Agriculture reliant on flooding was one of the reasons for the formation of certain ancient civilizations, such as the Ancient Egyptians near the Nile River. Floods can also bring nutrients into lakes and rivers, increasing biomass and supercharging local ecosystems. This may result in a temporary increase in local fish and bird populations. 

As floods are common events, creating flood models holds practical value. Teng ~\cite{Teng2017} presented a list of hydrodynamics models in which HEC-RAS was listed. The Hydrologic Engineering Center's River Analysis System (HEC-RAS), a program developed by the US Army Corp of Engineers, is capable of running 1D and 2D unsteady flow simulations and is ideal for flood prediction. HEC-RAS is also capable of performing sediment transport, movable boundary simulations and water quality analysis. Because of the program's support and versatility, HEC-RAS has been gaining popularity for flood simulations after the release of its version 5 ~\cite{HR19A} ~\cite{HR19B}. Various studies have been conducted using HEC-RAS since its early conception ~\cite{Davis1978} ~\cite{Devantier1993}. 

In 2016, the Egyptian government proposed road that was planned to run through the flood-prone Assiut Plateau near Durunka village. Inputting historical precipitation data into HEC-RAS, Ezz ~\cite{Ezz2016} was able to estimate the severity of flash floods in the region. The results from the study were able to identify locations where water could reach high velocities and thus, influence designers of the road to install protection and vents at those locations.

A 2017 study by Afshari ~\cite{Afshari2018} compared HEC-RAS to two other hydrodynamic modelling tools, AutoRoute and Height Above the Nearest Drainage (Hand). The programs were tested on several different terrains, including the Cedar River watershed in Iowa and an area near the Black Warrior River in Alabama, each with several different configurations. HEC-RAS proved to be capable of carrying out more complex flood simulations such as those involving levees and dams, and set the standard for accuracy for the three programs. Though, the simpler programs showed reasonably accurate prediction for low-complexity models and were more suited for situations where computational time was limited.

A recent study by Raman ~\cite{Raman2019} used HEC-RAS to simulate the effect of the Brumadinho dam break ~\cite{BBC19} with topographic dataset measurement based on NASA SRTM ~\cite{SRTM} and historical images archived by Google Earth. Comparison between HEC-RAS simulated results and satellite and photo images shows very similar inundation patterns for the mud flow as a result of the dam break. Significant life and property loss incurred due to the dam break could have been avoided if appropriate government policy and company practice took place ~\cite{Sant2019}, ~\cite{Jenn2019}. The environmental impact of mining dam construction and potential destruction should be evaluated for new dams planned in the US and around the globe ~\cite{WSJ2019}.

Eastern North Carolina has experienced three large floods in a span of two decades during Hurricane Floyd, Hurricane Matthew, and Hurricane Florence. Situated in this flooding hotbed is the Neuse River Floodplain in Craven County, NC. The area is relatively flat and swampy with a small cliff bordering its south side. As a result, overflow tends to spill out of the north side of the region into infrastructure and residential areas.

The results of Hurricane Florence are well-documented, with USGS obtaining high water marks and NOAA providing aerial imagery showing the extent of the flooding. Simulating flooding from Hurricane Florence provides for another opportunity to test capabilities of HEC-RAS.


\begin{figure}[htp]
    \centering
    \includegraphics[width=9cm]{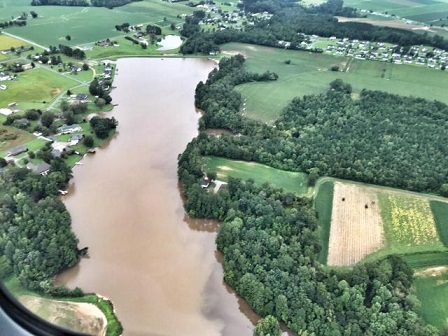}
    \caption{Neuse River, NC before flooding.}
    \label{fig:nri}
\end{figure}

  \section{Description of the event}
\begin{figure}[H]
		\begin{minipage}[b]{0.45\linewidth}
			\centering
			
			\includegraphics[width=9cm]{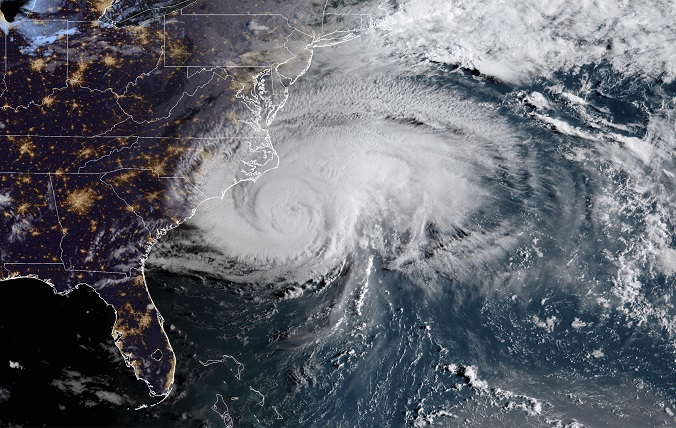}
			
			\caption{Artistic image of Hurricane Florence arriving at North Carolina.}
			\label{fig:hfai}
		\end{minipage}
		\hspace{1cm}
		\begin{minipage}[b]{0.45\linewidth}
			\centering
			
			\includegraphics[width=9cm]{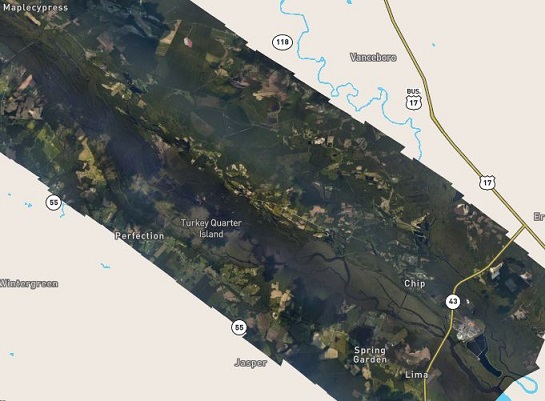}
			
			\caption{Satellite image of the freshwater flooding near Neuse River, North Carolina.}
			\label{fig:neuse_after}
		\end{minipage}
	\end{figure}

Hurricane Florence is a Category 4 Cape Verde hurricane with an extended  life span: August 31 - September 18, 2018. Originating from the west coast of Africa on August 30, 2018, it made landfall September 14, 7:15 EST in Wrightsville Beach, NC shown in Figure ~\ref{fig:hfai}, causing the most precipitation for a tropical cyclone recorded ~\cite{Connor2018} and ~\cite{Stewart2019} in North Carolina. There were 22 fatalities due to initial hit of the hurricane, and \$24 Billion in damages, most of which was due to freshwater flooding induced by the hurricane ~\cite{Fox19}. Large amounts of Rainfall and a marshy landscape near Neuse river caused significant freshwater spill on the north side of the river where the river bank is not high enough to contain the flooding. Figure ~\ref{fig:neuse_after} shows the area where this study is focused on.

\section{Theory of flood modelling}
  
  The full momentum equations solved by HEC-RAS are the 2D shallow water equations, in flux form
  
  \begin{eqnarray} \label{sweq}
  \pdv{\mathbf{ U}}{t}+\pdv{\mathbf{ F_x}}{x}+\pdv{\mathbf{ F_y}}{y} &=& \mathbf {S} \\
  \mathbf{U} &=& \begin{bmatrix}
                 h \\
                 hu \\
                 hv 
                 \end{bmatrix} \\
  \mathbf{F_x} &=& \begin{bmatrix}
                 hu \\
                 huu + 1/2 g h^2 \\
                 huv 
                 \end{bmatrix} \\
  \mathbf{F_y} &=& \begin{bmatrix}
                 hu \\
                 huv \\
                 hvv + 1/2 g h^2 \\
                 \end{bmatrix} \\
  \mathbf{S} &=& \begin{bmatrix} 
                 0 \\
                 - g h \pdv{z}{x} - c_f u \\
                 - g h \pdv{z}{y} - c_f v \\
                 \end{bmatrix} 
  \end{eqnarray} 

  where $\mathbf{U}$ is the state vector containing height and momentum of a fluid, $\mathbf{F_x}$ and $\mathbf{F_y}$ are the flux vectors
containing mass flux and momentum flux, and $\mathbf{S}$ is the source term describing bottom topography and friction. The bottom friction coefficient $c_f$ can be calculated from manning formula $c_f = \frac{n^2 g \sqrt{u^2+v^2}}{R}$ and is dependent on the Manning's coefficient $n$ ~\cite{HRRM2016}. HEC-RAS also adds Coriolis terms to the source term when instructed. 

\section{Modelling Discussion}

\subsection{Data collection}

 In this study, elevation data, spatial data and river station data are required inputs. Elevation data creates the shape of the terrain used for the flow area in the simulation. The topography is obtained with Light Detection and Ranging (LIDAR ~\cite{lidar}), a remote sensing method whereby a plane is flown over a certain area and determines its distance from the ground by scanning the terrain using pulsed lasers. The LIDAR digital elevation map used for this study has a 10m x 10m resolution and is provided by USGS ~\cite{USGS}. Spatial data for this study is provided by SpacialReference.org and gives HEC-RAS the location of the terrain, which allows the program to provide map overlays on top of the model. Elevation and spatial reference create the area in which the flow simulation can be done on. 

\begin{figure}[H]
		\begin{minipage}[b]{0.45\linewidth}
			\centering
			
			\includegraphics[width=9cm]{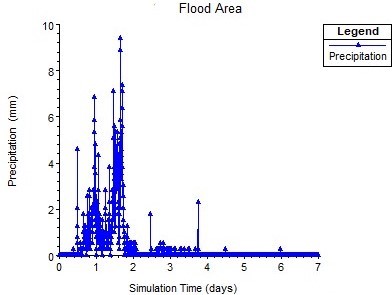}
			
			\caption{Precipitation data from USGS Highway 70 Station.}
			\label{fig:Rain_Data}
		\end{minipage}
		\hspace{1cm}
		\begin{minipage}[b]{0.45\linewidth}
			\centering
			
			\includegraphics[width=9cm]{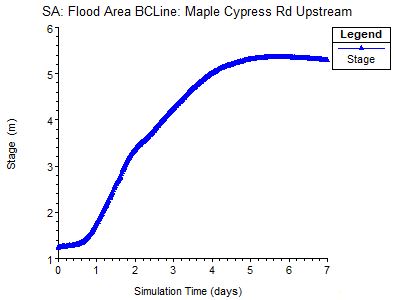}
			
			\caption{Gage height data from USGS Station near Fort Barnwell.}
			\label{fig:Fort_Barnwell}
		\end{minipage}
	\end{figure}

\begin{figure}[H]
        \centering
        \includegraphics[width=9cm]{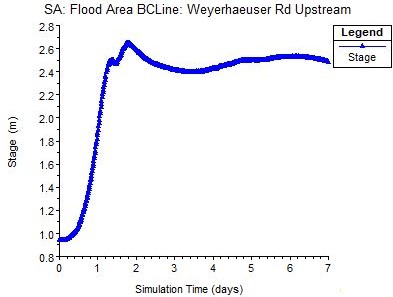}
        \caption{Gage height data from USGS Station near Streets Ferry.}
        \label{fig:Streets_Ferry}
    \end{figure}

For the data about a specific event such as a hurricane, river station data involves two important components: precipitation data and gage height data. During a hurricane, inland flooding is caused by heavy rainfall. Precipitation data was obtained from the USGS US Highway 70 Station and is shown in Figure~\ref{fig:Rain_Data} ~\cite{Highway70}. It gives the amount of rainfall every six minutes from September 13, 2018, 0:00:00 to September 19, 2018, 23:45:00. Precipitation also causes flooding upstream, which flows down into the flow area. Gage height data from the USGS Station near Fort Barnwell, shown in Figure~\ref{fig:Fort_Barnwell} ~\cite{Barn}, and the USGS Station near Streets Ferry, shown in Figure~\ref{fig:Streets_Ferry} ~\cite{StreetsFerry}, serve as the upstream boundary conditions. The gage height data has the same start and end time as the precipitation data, but has a time interval of 15 minutes instead of 6.

\subsection{Model setup}

Water flows into the flood area through the upstream and flows out through the downstream. If no downstream is set for a 2D flow area in HEC-RAS, the area's perimeter is treated as an infinitely high glass wall, allowing no water to exit. As this situation is not accurate for this model, downstream boundaries must be set with certain boundary conditions to determine how water flows out of the area. Water outflow behavior is assumed to have normal depth, whereby the slope of the bottom of a water channel is equal to the slope of the water surface. This occurs when the gravitational force on the water equals the frictional force at the bottom of the water channel. Normal depth is described by Manning's equations, and the Manning's roughness coefficient of friction used in this model for the flow area is 0.06, the default number in HEC-RAS. The friction slope of this model is set to 0.01, as the terrain is relatively flat. The floodplain is represented as a mesh with a 50m x 50m resolution resulting in 31082 cells over the flood simulation area. 

In computational fluid dynamics, calculating exact time derivatives with infinitely small time steps is not possible. So, picking the right size time step is imperative. In principle, a selected time step should be less than the time it takes for water to travel from one cell to the next. A time step that satisfies this condition is said to satisfy the Courant condition
$$C=\frac{V_w\Delta T}{\Delta X}\le 1$$
where $\mathbf{V_w}$ is the velocity of the water, $\mathbf{\Delta X}$ is the cell length and $\mathbf{\Delta T}$ is the time step. Too large of a time step can lead to the skipping over of cells, causing overestimations and instability. However, too small of a time step increases the number of calculations, which presents its own problems as well. Most notably, more calculations results in longer simulation run times and more computational approximations, which can also create model instability. Recent versions of HEC-RAS allows for the manual setting of the Courant number, which automatically finds a time step which satisfies the Courant condition. The Courant number used in this model is 0.9.

\subsection{Result}

\begin{figure}[H]
		\begin{minipage}[b]{0.45\linewidth}
			\centering
			
			\includegraphics[width=9cm]{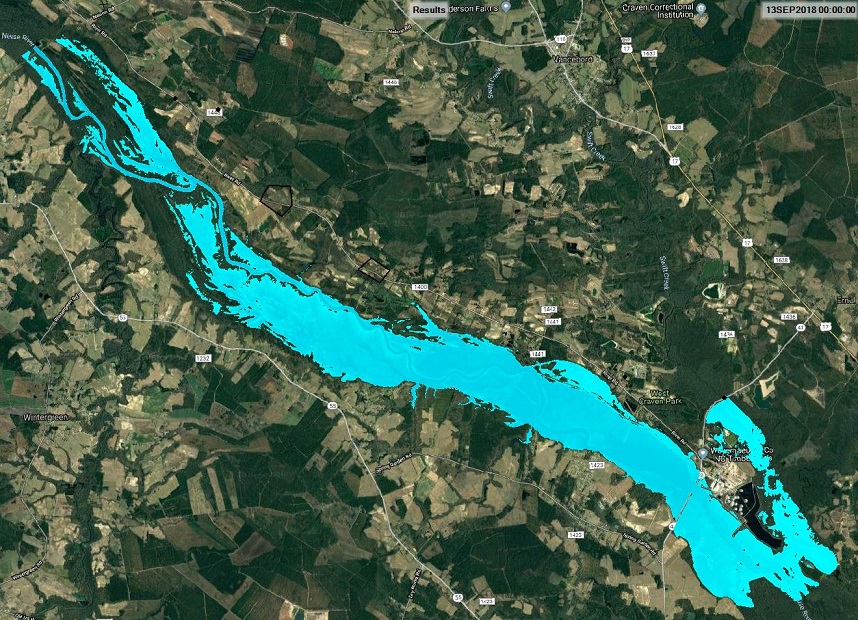}
			
			\caption{An image of the entire flood area on 13 September 2019, before the hurricane.}
			\label{fig:sim_overall_before}
		\end{minipage}
		\hspace{1cm}
		\begin{minipage}[b]{0.45\linewidth}
			\centering
			
			\includegraphics[width=9cm]{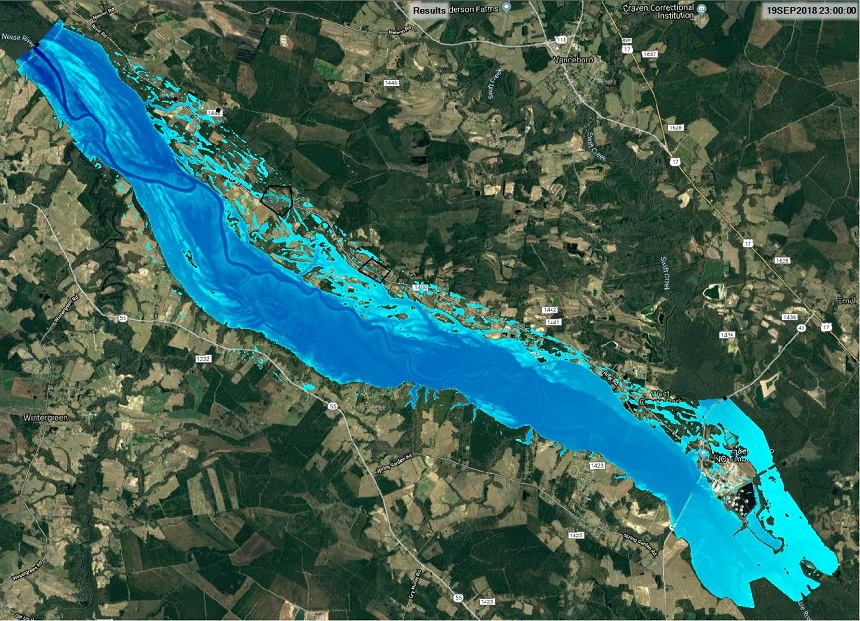}
			
			\caption{An image of the entire flood area on 19 September 2019, after the hurricane.}
			\label{fig:sim_overall_after}
		\end{minipage}
	\end{figure}

Figure~\ref{fig:sim_overall_before} and Figure~\ref{fig:sim_overall_after} are simulation pictures of the entire 2D flow. The Neuse River floodplain has a large amount of plant growth and foliage, making it difficult to observe how water is dispersed across the terrain from aerial images. Figure~\ref{fig:sim_overall_before} is an image of the simulation before the hurricane. It shows that the water is not only constrained to the center river strait. It is more spread out, with most of the area being submerged in water less than a meter high. This comes as no surprise as the area is known to be swampy. Figure~\ref{fig:sim_overall_after} is a simulated image of the inundation pattern after the rainfall from the hurricane, and shows water has extended beyond its normal confines.

\begin{figure}[H]
		\begin{minipage}[b]{0.45\linewidth}
			\centering
			
			\includegraphics[width=9cm]{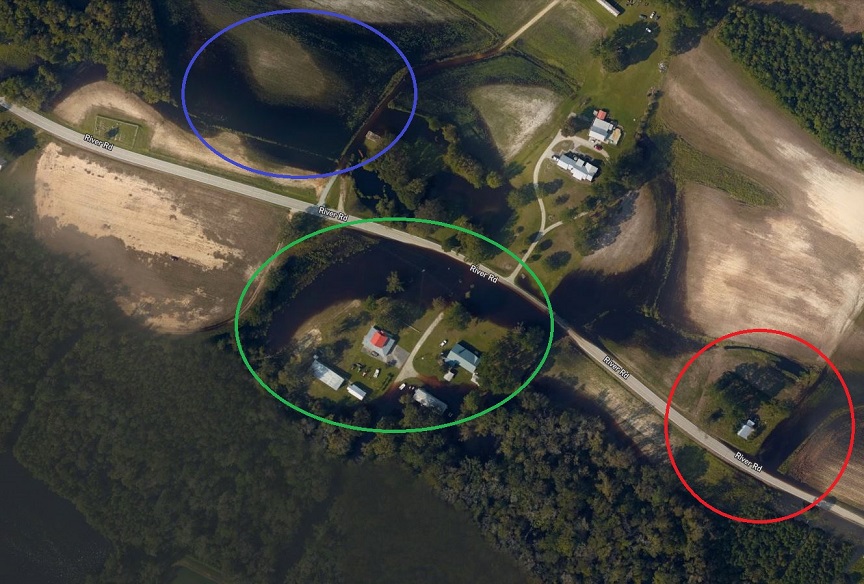}
			
			\caption{An actual flood image of Neuse Farms on 19 September 2019, after the hurricane.}
			\label{fig:neuse_farm_actual}
		\end{minipage}
		\hspace{1cm}
		\begin{minipage}[b]{0.45\linewidth}
			\centering
			
			\includegraphics[width=9cm]{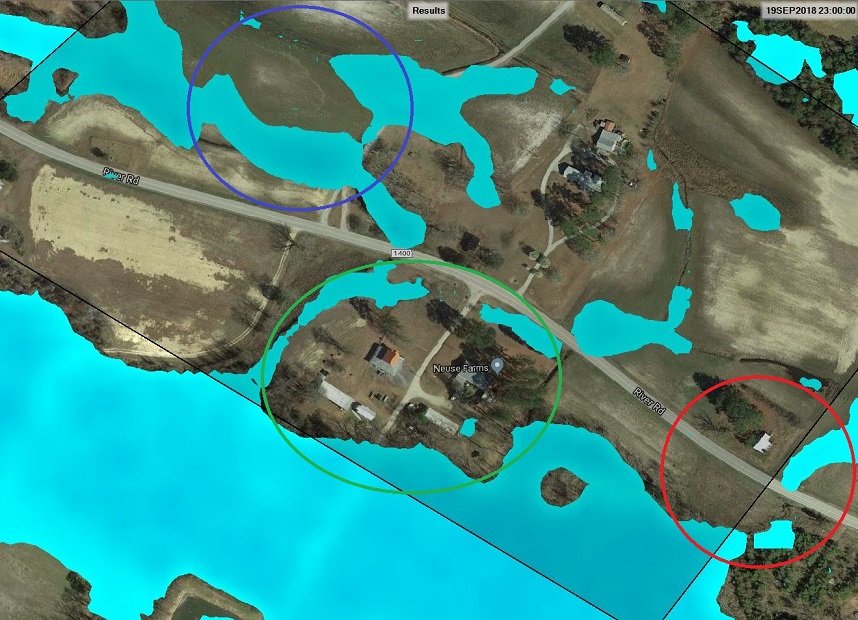}
			
			\caption{A simulated flood image of Neuse Farms on 19 September 2019, after the hurricane.}
			\label{fig:neuse_farm_sim}
		\end{minipage}
	\end{figure}

Figures~\ref{fig:neuse_farm_actual} and~\ref{fig:neuse_farm_sim} show the actual flood pattern and the simulated flood pattern of Neuse Farms respectively. Both of the areas circled in blue show the flood waters go around a hill, though there appears to be less water around the top of the image in the simulated flood. Both of the areas circled in green show the flood waters mostly go around the residential area. The actual image shows that one of the buildings is flooded whereas the simulated flood shows the waters just reaches that building. Both of the areas circled in red show flood waters reaching but not crossing the road.

\begin{figure}[H]
		\begin{minipage}[b]{0.45\linewidth}
			\centering
			
			\includegraphics[width=9cm]{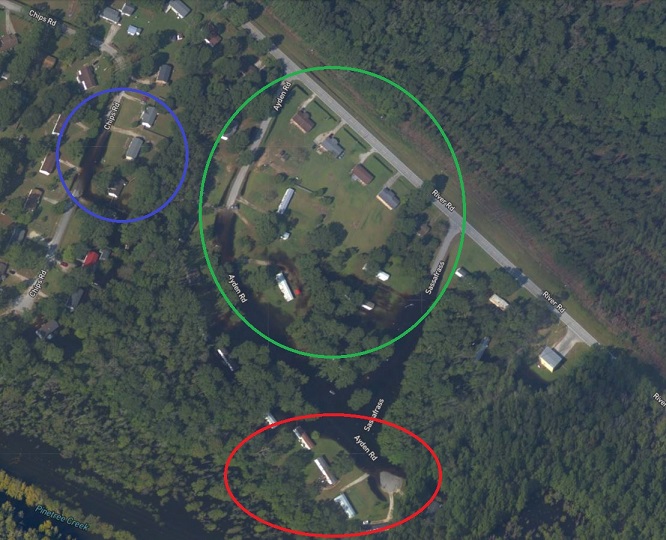}
			
			\caption{An actual flood image of Ayden Drive on 19 September 2019, after the hurricane.}
			\label{fig:ayden_dr_actual}
		\end{minipage}
		\hspace{1cm}
		\begin{minipage}[b]{0.45\linewidth}
			\centering
			
			\includegraphics[width=9cm]{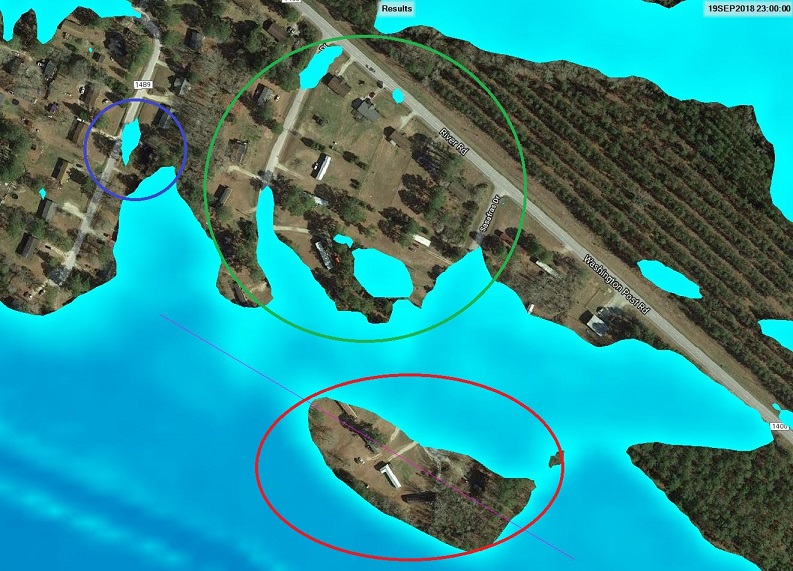}
			
			\caption{A simulated flood image of Ayden Drive on 19 September 2019, after the hurricane.}
			\label{fig:ayden_dr_sim}
		\end{minipage}
	\end{figure}

\begin{figure}[H]
        \centering
        \includegraphics[width=9cm]{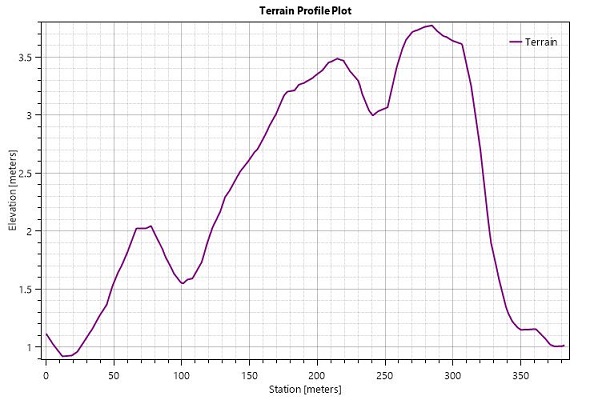}
        \caption{Terrain profile of highlighted blue and purple line in Figure~\ref{fig:ayden_dr_sim}.}
        \label{fig:ayden_dr_terrain}
    \end{figure}

Figures~\ref{fig:ayden_dr_actual} and~\ref{fig:ayden_dr_sim} show the actual flood pattern and simulated flood pattern of Ayden Drive respectively. Both of the areas circled in blue show the flood waters have reached Chips Road. Both of the areas circled in green show flood waters have reached and crept up Ayden and Sassafras Road. The simulation though, does have a small isolated patch of water of depth 10 centimeters near the intersection of Ayden and River Road that is not present in the actual flood image. Both the areas circled in red show a few homes surrounded by water but not flooded. Figure~\ref{fig:ayden_dr_terrain} shows why this phenomenon happens. The graph is an elevation profile of a cross section of terrain, specifically the cross section that is the blue and purple line in Figure~\ref{fig:ayden_dr_sim}. The terrain graph starts off at a low elevation of around 1 meter, climbs up to an elevation of over 3 meters towards the middle, then back down to around 1 meter. As shown in the red circle of Figure~\ref{fig:ayden_dr_sim}, the homes are built in the middle of the terrain cross section and are at a higher elevation than the surrounding land. As a result, it is possible for water to reach the surrounding land and not the homes. Water floods lower areas before it floods higher areas.

\begin{figure}[H]
		\begin{minipage}[b]{0.45\linewidth}
			\centering
			
			\includegraphics[width=9cm]{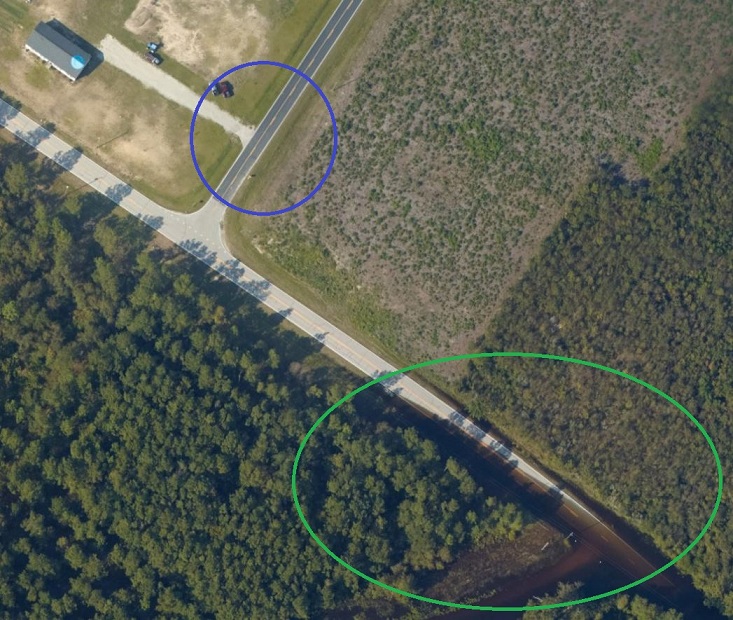}
			
			\caption{An actual flood image of a high water mark location on 19 September 2019, after the hurricane.}
			\label{fig:hwm_actual}
		\end{minipage}
		\hspace{1cm}
		\begin{minipage}[b]{0.45\linewidth}
			\centering
			
			\includegraphics[width=9cm]{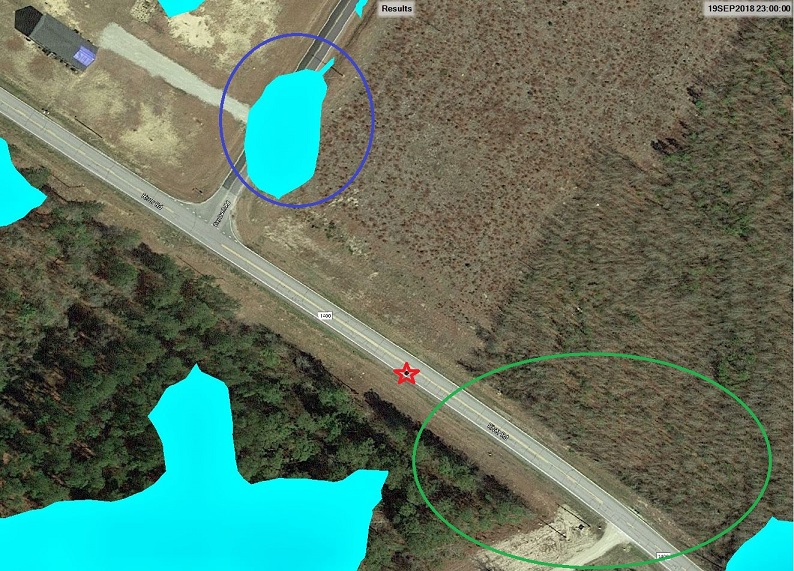}
			
			\caption{A simulated flood image of a high water mark location on 19 September 2019, after the hurricane.}
			\label{fig:hwm_sim}
		\end{minipage}
	\end{figure}

\begin{figure}[H]
        \centering
        \includegraphics[width=9cm]{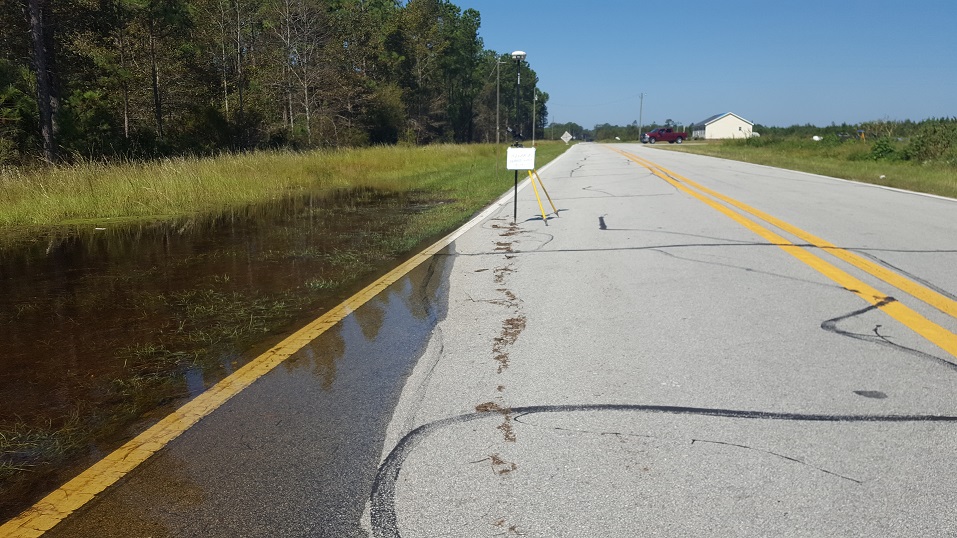}
        \caption{Street view of the debris line high water mark in Figure~\ref{fig:hwm_actual}.}
        \label{fig:hwm}
    \end{figure}

Figures~\ref{fig:hwm_actual} and~\ref{fig:hwm_sim} show the actual flood pattern and simulated flood pattern of the area near a high water mark ~\cite{HWM17}. Comparing the areas circled in blue in the images, the simulated flood imaged shows an approximately 19 centimeter deep puddle not present in the actual flood image. The area circled in green on the actual flood image shows the actual flood waters reached River Road whereas the same area is not flooded in the simulated flood image. Figure~\ref{fig:hwm} is a close-up image of the high water mark, measured by a debris line. The red star in Figure~\ref{fig:hwm_sim} corresponds with the location where the sign in Figure~\ref{fig:hwm} is placed. In this instance, the waters do not nearly get as far in the simulation compared to the actual flooding. Possible reasons include the resolution of the topography data is not high enough to resolve the road which is relatively flat and insufficient data of precipitation measurement over the area. Only one river station, USGS River Station near Highway 70, was near the flooding area and provided precipitation data measurements.

\section{Conclusions} 
Given topography and hydrological data from USGS, HEC-RAS solves a simplified version of the Navier-Stokes equations known as the shallow water equations to model flooding. In this study HEC-RAS was used to simulate flooding of the Neuse River during Hurricane Florence. Specific locations in the studied area were compared to actual observations. The agreement between simulated flooding result and observations is generally reasonable, with the difference between the observed cases and simulated results being around $\pm  20$ centimeters. The topography data is made available through USGS national map explorer. Three river stations near Neuse river provided precipitation and gage height data as input the HEC-RAS. 

Many factors could have limited the accuracy of the model. As there was no hydrographic data available for the downstream boundary conditions, the model used a normal depth condition at the outflow locations of the 2D flow area. Normal depth is more suited for constant discharge rates and is not ideal for simulating accelerating waters from flooding. Additionally, the precipitation data obtained for this simulation was obtained from a USGS center just outside the flood area. It also assumed the rainfall was the same throughout the whole 2D flow area, with no specific regions within the area having a higher rainfall rate than other regions in the area. More specific precipitation data, a higher resolution topography data, and the insertion of man made structures could potentially increase the accuracy of the flooding results ~\cite{Saksena15} ~\cite{Cost15}. As mentioned in ~\cite{SCIGage}, additional gauges were installed in North Carolina and Virginia to track the storm. But for the area near Neuse river, data were only partially available from the three river stations nearby. 

One critical feature that's missing in current HEC-RAS software is the effect of sustained wind on the motion of flood. Studies by ~\cite{SCIWind} indicates that flooding near New Bern which is exactly where this research is located, started before significant rainfall from the storm arrived since the water is driven upstream by sustained wind due to the storm. As explained by ~\cite{Hall19}, flooding can be slowed down and stalled due to strong wind from tropical storms. There is currently no way to introduce the wind effect in HEC-RAS. This is an area where HEC-RAS simulation result can be improved significantly.

Despite early warning, 34 people still died during hurrican Florence. And this is not a singular case ~\cite{SCIWind}. Early warning, hazard map, and local government assistance are still not enough to prevent the loss of lives in events like this. Public education and awareness of the destructive effect of tropical storm often fall short. As climate gains increased variability, tropical cylcone activities will most likely rise including increased sea level rise, cyclone rainfall rates, overall intensity of tropical cyclones, and number of tropical cyclones over category 4 and 5 ~\cite{BM1}, ~\cite{BM2}, ~\cite{GFDL19}. Both predicative and preventative measures to reduce destructive effect of hurricanes will continue to be a focus of scientific, technical and political effort.


\end {document}